\documentclass[12pt]{article}
\usepackage{amsmath,amssymb}
\usepackage{bm}
\usepackage{color}
\usepackage{graphicx}
\usepackage{cite}

\begin{document}
\title{
\begin{flushright}
\ \\*[-80pt]
\begin{minipage}{0.2\linewidth}
\normalsize
HUPD1705 \\*[50pt]
\end{minipage}
\end{flushright}
{\Large \bf
Phenomenological Aspects of Possible Vacua of a Neutrino Flavor Model
\\*[20pt]}}

\author{
\centerline{
Takuya~Morozumi$^{1,2,}$\footnote{E-mail address: morozumi@hiroshima-u.ac.jp},
~Hideaki~Okane$^{1,}$\footnote{E-mail address: hideaki-ookane@hiroshima-u.ac.jp},
~Hiroki~Sakamoto$^{1,}$\footnote{E-mail address: h-sakamoto@hiroshima-u.ac.jp},} \\
\centerline{Yusuke~Shimizu$^{1,}$\footnote{E-mail address: yu-shimizu@hiroshima-u.ac.jp},
~Kenta~Takagi$^{1,}$\footnote{E-mail address: takagi-kenta@hiroshima-u.ac.jp},~and
~Hiroyuki~Umeeda$^{3,}$\footnote{E-mail address: umeeda@riko.shimane-u.ac.jp}}
\\*[20pt]
\centerline{
\begin{minipage}{\linewidth}
\begin{center}
$^1${\it \normalsize
Graduate~School~of~Science,~Hiroshima~University, \\
Higashi-Hiroshima~739-8526,~Japan} \\*[5pt]
$^2${\it \normalsize
Core~of~Research~for~the~Energetic~Universe,~Hiroshima~University, \\
Higashi-Hiroshima~739-8526,~Japan} \\*[5pt]
$^3${\it \normalsize
Graduate~School~of~Science~and~Engineering,~Shimane~University, \\
Matsue 690-8504, Japan
}
\end{center}
\end{minipage}}
\\*[50pt]}

\date{
\centerline{\small \bf Abstract}
\begin{minipage}{0.9\linewidth}
\medskip
\medskip
\small
~~We discuss a supersymmetric model with discrete flavor symmetry $A_4\times Z_3$.
The additional scalar fields which contribute masses of leptons in the Yukawa terms are introduced in this model.
We analyze their scalar potential and find that they have various vacuum structures.
We show the relations among 24 different vacua and classify them into two types.
We derive expressions of the lepton mixing angles, Dirac CP violating phase and Majorana phases for the two types.
The model parameters which are allowed by the experimental data of the lepton mixing angles are different for each type.
We also study the constraints on the model parameters which are related to Majorana phases.
The different allowed regions of the model parameters for the two types are shown numerically for a given region of two combinations of the CP violating phases.
\end{minipage}
}

\begin{titlepage}
\maketitle
\thispagestyle{empty}
\end{titlepage}

\section{Introduction}
Although all the elementary particles in the standard model (SM) have now been discovered, with the discovery of the Higgs boson,
there still exist phenomena which cannot be explained in the framework of the SM.
One of these is the neutrino oscillation phenomenon,
which implies two non-zero neutrino mass squared differences and two large lepton mixing angles.
In order to explain this, many authors propose a neutrino flavor model with non-Abelian discrete flavor symmetry
in the lepton sector (for reviews see~\cite{Ishimori:2010au,Ishimori:2012zz,Ishimori:2013woa,King:2014nza}).
Even before the discovery of the non-zero $\theta _{13}$~\cite{An:2012eh,Ahn:2012nd,Abe:2014lus},
a few authors suggested a tiny mixing angle $\theta _{13}$ based on  non-Abelian discrete flavor symmetry~\cite{Shimizu:2011xg}.
Recent results from the T2K and NO$\nu $A experiments~\cite{Abe:2013hdq,Adamson:2016tbq} imply CP violation through the Dirac CP phase.
They studied electron neutrino appearance in a muon neutrino beam.
The Majorana phases are also sources of the CP violating phases if neutrinos are Majorana particles.
The KamLAND-Zen experiment~\cite{KamLAND-Zen:2016pfg} is searching for neutrinoless double beta ($0\nu \beta \beta $) decay to check the Majorana nature of neutrinos.
Therefore, it is important to predict not only mixing angles but also CP phases with the non-Abelian discrete flavor model.

The non-Abelian discrete flavor symmetry can easily explain large lepton mixing angles,
e. g. tri-bimaximal mixing (TBM)~\cite{Harrison:2002er,Harrison:2002kp}, which is a simple framework for the lepton mixing angles.
Indeed, Altarelli and Feruglio (AF) proposed a simple flavor model and predicted TBM by using $A_4$ discrete flavor symmetry~\cite{Altarelli:2005yp,Altarelli:2005yx}.
They introduced $SU(2)$ gauge singlet scalar fields, so-called ``flavons'', and derived the TBM in the lepton sector.
The non-zero $\theta_{13}$ can be realized by another $A_4$ non-trivial singlet flavon~\cite{Shimizu:2011xg} in addition to the flavons introduced by AF.
The origin of non-vanishing $\theta_{13}$ is related to a new contribution to the mass matrices.
Matrices which have the same structure as that in Ref.~\cite{Shimizu:2011xg} also appear in extra-dimensional models with the $S_3$ and $S_4$ flavor symmetries~\cite{Haba:2006dz,Ishimori:2010fs}.
The $\Delta (27)$ model also includes these matrices~\cite{Grimus:2008tt}.

In this paper, we study phenomenological aspects of a supersymmetric model with $A_4\times Z_3$ symmetries.
The three generations of the left-handed leptons are expressed as the $A_4$ triplet, $l=(l_e,l_\mu,l_\tau )$,
while the right-handed charged leptons $e_R$, $\mu_R$, and $\tau_R$ are $A_4$ singlets denoted as ${\bf 1}$, ${\bf 1''}$, and ${\bf 1'}$ respectively.
Three right-handed neutrinos are also described as the triplet of $A_4$.
We introduce the $SU(2)$ gauge singlet flavons of $A_4$ triplets, $\phi_T=(\phi _{T1},\phi _{T2},\phi _{T3})$ and $\phi_S=(\phi _{S1},\phi _{S2},\phi _{S3})$.
In addition, $\xi $ and $\xi '$ are also introduced as the $SU(2)$ gauge singlet flavons with the two kinds of singlet representations of $A_4$, ${\bf 1}$ and ${\bf 1'}$ respectively.

We focus on the vacuum structure of the flavor model.
The scalar sectors of this model consist of many flavons in addition to the SM Higgs boson.
Then, we analyze the scalar potential and show the 24 different sets of VEVs
which come from 24 combinations of 4 (6) possible VEVs of the flavon $\phi_T$ ($\phi_S$).
The 24 different vacua are classified into two types which are not related to each other under the transformations $A_4$.
Therefore, we expect that the two types of vacua have different expressions for the physical observables in terms of the model parameters such as Yukawa couplings.
We ask the following question: whether these different vacua are physically distinct from each other.
The purpose of this paper is to clarify the differences and relations among the VEVs and their physical consequences.
In particular, we investigate the mixing angles, CP violating phase, and effective mass for neutrinoless double beta ($0\nu\beta\beta$) decay.

This paper is organized as follows.
In Section~\ref{sec:model}, we introduce the supersymmetric model with $A_4\times Z_3$ symmetry.
In Section~\ref{sec:2types}, we study the classification of vacua and derive the formulae for the mixing angles and CP phases.
In Section~\ref{sec:num}, we discuss the phenomenological aspects for mixing angles and CP violating phases.
The numerical analyses for the effective mass of $0\nu\beta\beta$ decay are presented.
Section~\ref{sec:con} is devoted to a summary.
In Appendix~\ref{sec:multiplication-rule}, we show the multiplication rule of the $A_4$ group.

\section{Supersymmetric Model with $A_4\times Z_3$ Symmetry}
\label{sec:model}
In this section, we introduce a supersymmetric model with $A_4\times Z_3$ symmetry.
We analyze the scalar potential and derive the mass matrices of the lepton sector.
\subsection{Model}
We introduce three heavy right-handed Majorana neutrinos.
The leptons and scalars in our model are listed in Table~\ref{tb:fields}.
\begin{table}[b]
	\caption{\small{
	The representations of $SU(2)_L$ and $A_4$, and the charge assignment of $Z_3$ and $U(1)_R$ for leptons and scalars:
	~$l_{e,\mu,\tau}$, $\{e,\mu,\tau\}_R$, $\{\nu_e,\nu_\mu,\mu_\tau\}_R$, and $h_{u,d}$ denote left-handed leptons,
	right-handed charged leptons, right-handed neutrinos, and Higgs fields, respectively.
	The other scalars are gauge singlet flavons and denoted as $\phi_T$, $\phi_S$, $\xi$, and $\xi'$.
	$\omega$ is the $Z_3$ charge and stands for $e^{2\pi i/3}$.}}
	\hspace{-0.5cm}
 	\begin{tabular}{|c||c|c|c|c|c||c|c|c|c|c|} \hline

		&$l=	\begin{pmatrix}
			l_e\\
			l_\mu\\
			l_\tau
		\end{pmatrix}$
		&$e_R$
		&$\mu_R$
		&$\tau_R$
		&$\nu_R=\begin{pmatrix}
			\nu_{eR}\\
			\nu_{\mu R}\\
			\nu_{\tau R}
		\end{pmatrix}$
		&$h_{u,d}$
		&$\phi_T=\begin{pmatrix}
			\phi_{T1}\\
			\phi_{T2}\\
			\phi_{T3}
		\end{pmatrix}$
		&$\phi_S=\begin{pmatrix}
			\phi_{S1}\\
			\phi_{S2}\\
			\phi_{S3}
		\end{pmatrix}$
		&$\xi$
		&$\xi'$\\ \hline \hline

		$SU(2)_L$&2&1&1&1&1&2&1&1&1&1\\

		$A_4$&3&1&$1''$&$1'$&3&1&3&3&1&$1'$\\

		$Z_3$&$\omega$&$\omega^2$&$\omega^2$&$\omega^2$&$\omega^2$&1&1&$\omega^2$&$\omega^2$&$\omega^2$\\

		$U(1)_R$&1&1&1&1&1&0&0&0&0&0\\ \hline
		
	\end{tabular}\label{tb:fields}
\end{table}
The superpotential of Yukawa interactions is
\begin{align}
	w_Y=w_l+w_D+w_R,\label{eq:wY}
\end{align}
where $w_l,w_D$ and $w_R$ are Yukawa interactions for charged lepton, Dirac neutrino and Majorana neutrino sectors respectively:
\begin{align}
	&w_l=
	y_e(\phi_Tl)_{\bf 1}e_Rh_d/\Lambda+
	y_\mu(\phi_Tl)_{\bf 1'}\mu_Rh_d/\Lambda+
	y_\tau(\phi_Tl)_{\bf 1''}\tau_Rh_d/\Lambda+h.c., \label{eq:wl} \\
	&w_D=y_D(l\nu_R)_{\bf 1}h_u+h.c., \label{eq:wD} \\
	&w_R=y_{\phi_S}\phi_S(\nu_R\nu_R)_{\bf 3}+
	y_\xi\xi(\nu_R\nu_R)_{\bf 1}+
	y_{\xi'}\xi'(\nu_R\nu_R)_{\bf 1''}+h.c., \label{eq:wR}
\end{align}
where the lower indices denote $A_4$ representations.
Moreover, the $y$'s and $\Lambda$ denote the Yukawa coupling constants and cut-off scale respectively.
The multiplication rule for $A_4$ representations is shown in Appendix~\ref{sec:multiplication-rule}.

In order to obtain the mass matrices of these leptons, we analyze the following superpotential of the scalar fields:
\begin{align}
	w_d\equiv w_d^T+w_d^S, \label{eq:wd}
\end{align}
where
\begin{align}
	&w_d^T=-M(\phi_0^T\phi_T)_{\bf 1}+g\phi_0^T(\phi_T\phi_T)_{\bf 3}, \label{eq:wdT}  \\
	&w_d^S=
	g_1\phi_0^S(\phi_S\phi_S)_{\bf 3}+
	g_2(\phi_0^S\phi_S)_{\bf 1}\xi+g_2'(\phi_0^S\phi_S)_{\bf 1''}\xi '+
	g_3(\phi_S\phi_S)_{\bf 1}\xi_0-
	g_4\xi_0\xi\xi \label{eq:wdS}.
\end{align}
We have introduced the additional $SU(2)$ gauge singlet fields, $\phi_0^T$, $\phi_0^S$ and $\xi_0$, which are called {\it ``driving fields''}.
The charge assignments of these fields are summarized in Table \ref{tb:driving}.
\begin{table}[h!]
	\caption{The driving fields and their representations and charge assignment.}
	\centering
	\begin{tabular}{|c||c|c|c|} \hline
		 & $\phi_0^T=\begin{pmatrix}
		 	\phi_{01}^T\\
		 	\phi_{02}^T\\
		 	\phi_{03}^T
		\end{pmatrix}$
		 & $\phi_0^S=\begin{pmatrix}
		 	\phi_{01}^S\\
		 	\phi_{02}^S\\
		 	\phi_{03}^S
		\end{pmatrix}$
		 & $\xi_0$ \\ \hline \hline

		SU(2) & $1$ & $1$ & $1$ \\
		$A_4$ & $3$ & $3$ & $1$ \\
		$Z_3$ &  $1$ & $\omega^2$ & $\omega^2$ \\
		$U(1)_R$ &  $2$ & $2$ & $2$ \\ \hline
	\end{tabular}\label{tb:driving}
\end{table}

\subsection{Potential Analysis}
In this subsection, we derive the VEVs for the scalar fields $\phi_T,\phi_S,\xi,\xi',\phi^T_0,\phi^S_0,\xi_0$.
One can derive the scalar potential from the superpotentials in Eqs.~\eqref{eq:wdT} and \eqref{eq:wdS} as
\begin{align}
	V=V_T+V_S,
\end{align}
where
\begin{align}
	V_T=\sum_{X}\left|\frac{\partial w_d^T}{\partial X}\right|^2
	=&\left| -M\phi_{T1}+\frac{2}{3}g(\phi_{T1}^2-\phi_{T2}\phi_{T3})\right|^2 \nonumber \\
	+&\left| -M\phi_{T3}+\frac{2}{3}g(\phi_{T2}^2-\phi_{T3}\phi_{T1})\right|^2 \nonumber \\
	+&\left| -M\phi_{T2}+\frac{2}{3}g(\phi_{T3}^2-\phi_{T1}\phi_{T2})\right|^2 \nonumber \\
	+&\left| -M\phi_{01}^T+\frac{2}{3}g(2\phi_{01}^T\phi_{T1}-\phi_{03}^T\phi_{T2}-\phi_{02}^T\phi_{T3})\right|^2 \nonumber \\
	+&\left| -M\phi_{03}^T+\frac{2}{3}g(2\phi_{02}^T\phi_{T2}-\phi_{01}^T\phi_{T3}-\phi_{03}^T\phi_{T1})\right|^2 \nonumber \\
	+&\left| -M\phi_{02}^T+\frac{2}{3}g(2\phi_{03}^T\phi_{T3}-\phi_{02}^T\phi_{T1}-\phi_{01}^T\phi_{T2})\right|^2  \label{eq:VT},
\end{align}
and
\begin{align}
	V_S
	=\sum_{Y}\left|\frac{\partial w_d^S}{\partial Y}\right|^2
	=&\left|\frac{2}{3}g_1(\phi_{S1}^2-\phi_{S2}\phi_{S3})-g_2\phi_{S1}\xi+g_2'\phi_{S3}\xi'\right|^2 \nonumber \\
	+&\left|\frac{2}{3}g_1(\phi_{S2}^2-\phi_{S3}\phi_{S1})-g_2\phi_{S3}\xi+g_2'\phi_{S2}\xi'\right|^2 \nonumber \\
	+&\left|\frac{2}{3}g_1(\phi_{S3}^2-\phi_{S1}\phi_{S2})-g_2\phi_{S2}\xi+g_2'\phi_{S1}\xi'\right|^2 \nonumber \\
	+&\left|\frac{2}{3}g_1(2\phi_{01}^S\phi_{S1}-\phi_{03}^S\phi_{S2}-\phi_{02}^S\phi_{S3})
	-g_2\phi_{01}^S\xi+g_2'\phi_{03}^S\xi'+2g_3\phi_{S1}\xi_0\right|^2 \nonumber \\
	+&\left|\frac{2}{3}g_1(2\phi_{02}^S\phi_{S2}-\phi_{01}^S\phi_{S3}-\phi_{03}^S\phi_{S1})
	-g_2\phi_{03}^S\xi+g_2'\phi_{02}^S\xi'+2g_3\phi_{S3}\xi_0\right|^2 \nonumber \\
	+&\left|\frac{2}{3}g_1(2\phi_{03}^S\phi_{S3}-\phi_{02}^S\phi_{S1}-\phi_{01}^S\phi_{S2})
	-g_2\phi_{02}^S\xi+g_2'\phi_{01}^S\xi'+2g_3\phi_{S2}\xi_0\right|^2 \nonumber \\
	+&\left|-g_2(\phi_{01}^S\phi_{S1}+\phi_{03}^S\phi_{S2}+\phi_{02}^S\phi_{S3})-2g_4\xi\xi_0\right|^2 \nonumber \\
	+&\left|g_2'(\phi_{02}^S\phi_{S2}+\phi_{01}^S\phi_{S3}+\phi_{03}^S\phi_{S1})\right|^2 \nonumber \\
	+&\left|g_3(\phi_{s1}^2+2\phi_{S2}\phi_{S3})-g_4\xi^2\right|^2  \label{eq:VS}.
\end{align}
The sum for $X,Y$ runs over all the scalar fields:
$$
X=\{\phi_{T1}, \phi_{T2}, \phi_{T3}, \phi_{01}^T, \phi_{02}^T, \phi_{03}^T\}\ , \quad
Y=\{\phi_{S1}, \phi_{S2}, \phi_{S3}, \phi_{01}^S, \phi_{02}^S, \phi_{03}^S, \xi, \xi ', \xi_0\}.
$$
The scalar potential $V$ is minimized at $V=V_T=V_S=0$.
There are several solutions for the minimization condition.
We obtain  sets of solutions denoted as $\eta_m$ and $\lambda_n^\pm$ ($m=1$-$4$, $n=1$-$3$),
where $\eta_m$ and $\lambda_n^\pm$ are the solutions of $V_T=0$ and $V_S=0$ respectively.
Hereafter, we call them the set of VEV alignments and show them explicitly as follows:
\begin{align}
	&\eta_1\equiv\left\{\langle\phi_T\rangle=v_T\begin{pmatrix}1\\0\\0\end{pmatrix},\ 
	\langle\phi_0^T\rangle=\begin{pmatrix}0\\0\\0\end{pmatrix}\right\},\label{eq:eta1}\\
	&\eta_2\equiv\left\{\langle\phi_T\rangle=\frac{v_T}{3}\begin{pmatrix}-1\\2\\2\end{pmatrix},\ 
	\langle\phi_0^T\rangle=\begin{pmatrix}0\\0\\0\end{pmatrix}\right\},\label{eq:eta2}\\
	&\eta_3\equiv\left\{\langle\phi_T\rangle=\frac{v_T}{3}\begin{pmatrix}-1\\2\omega\\2\omega^2\end{pmatrix},\ 
	\langle\phi_0^T\rangle=\begin{pmatrix}0\\0\\0\end{pmatrix}\right\},\label{eq:eta3}\\
	&\eta_4\equiv\left\{\langle\phi_T\rangle=\frac{v_T}{3}\begin{pmatrix}-1\\2\omega^2\\2\omega\end{pmatrix},\ 
	\langle\phi_0^T\rangle=\begin{pmatrix}0\\0\\0\end{pmatrix}\right\},\label{eq:eta4}\\
	&\lambda_1^\pm\equiv\left\{\langle\phi_S\rangle=\pm v_S\begin{pmatrix}1\\1\\1\end{pmatrix},\ 
	\langle\xi'\rangle=u',\ \langle\phi_0^S\rangle=\begin{pmatrix}0\\0\\0\end{pmatrix}\right\},\label{eq:lam1}\\
	&\lambda_2^\pm\equiv\left\{\langle\phi_S\rangle=\pm v_S\begin{pmatrix}1\\\omega\\\omega^2\end{pmatrix},\ 
	\langle\xi'\rangle=\omega u',\ \langle\phi_0^S\rangle=\begin{pmatrix}0\\0\\0\end{pmatrix}\right\},\label{eq:lam2}\\
	&\lambda_3^\pm\equiv\left\{\langle\phi_S\rangle=\pm v_S\begin{pmatrix}1\\\omega^2\\\omega\end{pmatrix},\ 
	\langle\xi'\rangle=\omega^2u',\ \langle\phi_0^S\rangle=\begin{pmatrix}0\\0\\0\end{pmatrix}\right\},\label{eq:lam3}
\end{align}
where $v_T=\frac{3M}{2g}$, $v_S=\sqrt{\frac{g_4}{3g_3}}u$, $u'=\frac{g_2}{g_2'}u$ and $u$ is the VEV of $\xi$, $\langle\xi\rangle=u$
\footnote{
	There are still other solutions for $V=0$, including the trivial solution which makes all the VEVs vanish.
	It leads to the vanishing of all the lepton masses and mixing angles.
	In addition to the trivial solution, there are solutions with  non-zero VEVs of the driving fields.
	This case leads to the breakdown of $U(1)_R$ symmetry.
	In this paper, we only discuss the vacua where  $U(1)_R$ symmetry is conserved.}.
The superscript of $\lambda^\pm$ denotes the overall sign of the VEV $\langle\phi_S\rangle$.
In total, we obtain 24 sets of vacua, since there are four sets of alignment for $\eta_m$ and six sets for $\lambda^\pm_n$.
\subsection{Mass Matrix for Charged Leptons and Neutrinos}
We derive charged lepton mass matrices and neutrino mass matrices from the Yukawa interactions in Eqs.~(\ref{eq:wl}),(\ref{eq:wD}), and (\ref{eq:wR}).
These matrices are expressed in various forms corresponding to the VEV alignments.
The charged lepton mass matrices ${M_l}^{(m)}$ for Eqs.~\eqref{eq:eta1}-\eqref{eq:eta4} are
\begin{align}
	&{M_l}^{(1)}=\frac{v_dv_T}{\Lambda}
	\begin{pmatrix}
		y_e&0&0\\
		0&y_\mu&0\\
		0&0&y_\tau
	\end{pmatrix},
	\label{eq:Ml1} \\[0.3cm]
	&{M_l}^{(2)}=\frac{v_dv_T}{3\Lambda}
	\begin{pmatrix}
		-y_e&2y_\mu&2y_\tau\\
		2y_e&-y_\mu&2y_\tau\\
		2y_e&2y_\mu&-y_\tau
	\end{pmatrix}
	=S{M_l}^{(1)}, \label{eq:Ml2} \\[0.3cm]
	&{M_l}^{(3)}=\frac{v_dv_T}{3\Lambda}
	\begin{pmatrix}
		-y_e&2\omega y_\mu&2\omega^2y_\tau\\
		2\omega^2y_e&-y_\mu&2\omega y_\tau\\
		2\omega y_e&2\omega^2y_\mu&-y_\tau
	\end{pmatrix}
	=T^\dag ST{M_l}^{(1)}, \label{eq:Ml3} \\[0.3cm]
	&{M_l}^{(4)}=\frac{v_dv_T}{3\Lambda}
	\begin{pmatrix}
		-y_e&2\omega^2y_\mu&2\omega y_\tau\\
		2\omega y_e&-y_\mu&2\omega^2y_\tau\\
		2\omega^2y_e&2\omega y_\mu&-y_\tau
	\end{pmatrix}
	=TST^\dag{M_l}^{(1)}, \label{eq:Ml4}
\end{align}
respectively, where the matrices $S$ and $T$ are
\begin{align}
	S=\frac{1}{3}\begin{pmatrix}
		-1&2&2\\
		2&-1&2\\
		2&2&-1
	\end{pmatrix}\ ,\quad
	T=\begin{pmatrix}
		1&0&0\\
		0&\omega&0\\
		0&0&\omega^2
	\end{pmatrix}.
\label{eq:ST}
\end{align}
The Dirac mass matrix for neutrinos obtained from Eq.~\eqref{eq:wD} is
\begin{align}
	M_D=y_Dv_u\begin{pmatrix}
		1&0&0\\
		0&0&1\\
		0&1&0
	\end{pmatrix}.
\label{eq:MD}
\end{align}
It is noted that the Dirac mass matrix is determined independently of the VEV alignments.
The Majorana mass matrices ${M_R^{(n)}}^\pm$ for the corresponding set of solutions Eqs.~\eqref{eq:lam1},\eqref{eq:lam2},\eqref{eq:lam3} are given as follows:
\begin{align}
	{M_R^{(1)}}^\pm&=
	\pm\frac{1}{3}y_{{\phi_S}}v_S\begin{pmatrix}
		2&-1&-1\\
		-1&2&-1\\
		-1&-1&2
	\end{pmatrix}+y_\xi u\begin{pmatrix}
		1&0&0\\
		0&0&1\\
		0&1&0
	\end{pmatrix}+y_{\xi'}u'\begin{pmatrix}
		0&0&1\\
		0&1&0\\
		1&0&0
	\end{pmatrix}, \label{eq:MR1} \\
	{M_R^{(2)}}^\pm&=
	\pm\frac{1}{3}y_{{\phi_S}}v_S\begin{pmatrix}
		2&-\omega^2&-\omega\\
		-\omega^2&2\omega&-1\\
		-\omega&-1&2\omega^2
	\end{pmatrix}+y_\xi u\begin{pmatrix}
		1&0&0\\
		0&0&1\\
		0&1&0
	\end{pmatrix}+\omega y_{\xi'}u'\begin{pmatrix}
		0&0&1\\
		0&1&0\\
		1&0&0
	\end{pmatrix}\nonumber\\
	&=T^\dag{M_R^{(1)}}^\pm T^\dag, \label{eq:MR2} \\
	{M_R^{(3)}}^\pm&=
	\pm\frac{1}{3}y_{{\phi_S}}v_S\begin{pmatrix}
		2&-\omega&-\omega^2\\
		-\omega&2\omega^2&-1\\
		-\omega^2&-1&2\omega
	\end{pmatrix}+y_\xi u\begin{pmatrix}
		1&0&0\\
		0&0&1\\
		0&1&0
	\end{pmatrix}+\omega^2y_{\xi'}u'\begin{pmatrix}
		0&0&1\\
		0&1&0\\
		1&0&0
	\end{pmatrix}\nonumber\\
	&=T{M_R^{(1)}}^\pm T. \label{eq:MR3}
\end{align}

In order to generate the light neutrino mass matrices, we adopt the seesaw mechanism~\cite{Minkowski,Yanagida:1979as,GellMann:1980vs}.
The effective neutrino mass matrices are given by the well-known formula, $M_\nu=-M_DM_R^{-1}M_D^\mathrm{T}$,
through the seesaw mechanism.
We obtain the 6 different effective neutrino mass matrices from Eqs.~\eqref{eq:MD}-\eqref{eq:MR3} as follows:
\begin{align}
	&{M_\nu^{(1)}}^\pm=\pm a\begin{pmatrix}
		1&0&0\\
		0&1&0\\
		0&0&1
	\end{pmatrix}+b^\pm\begin{pmatrix}
		1&1&1\\
		1&1&1\\
		1&1&1
	\end{pmatrix}+c\begin{pmatrix}
		1&0&0\\
		0&0&1\\
		0&1&0
	\end{pmatrix}+d\begin{pmatrix}
		0&0&1\\
		0&1&0\\
		1&0&0
	\end{pmatrix}, \label{eq:Mv1}\\
	&{M_\nu^{(2)}}^\pm=T^\dag{M_\nu^{(1)}}^\pm T^\dag, \label{eq:Mv2}\\
	&{M_\nu^{(3)}}^\pm=T{M_\nu^{(1)}}^\pm T, \label{eq:Mv3}
\end{align}
where
\begin{align}
	&a=ky_{{\phi_S}}v_S, \nonumber\\
	&c=k(y_{\xi'}u'-y_\xi u), \nonumber\\
	&d=ky_{\xi'}u',  \nonumber\\
	&b^\pm=\mp\frac a3+\frac{a^2}{2d-c}\left(\frac13-\frac{d^2}{a^2}\right), \nonumber\\
	&k=\frac{{y_D}^2{v_u}^2}{y_\xi^2u^2+y_{\xi'}^2{u'}^2-(y_{\phi_S}^2v_S^2+y_\xi uy_\xi'u')}. \nonumber
\end{align}
\section{Classification of Vacua and PMNS Mixing Matrix}\label{sec:2types}
In this section, we classify the 24 different vacua and derive the lepton mixing matrix $U_\mathrm{PMNS}$, called the {\it Pontecorvo-Maki-Nakagawa-Sakata} (PMNS) mixing matrix.
In order to classify the vacua, we discuss the relations among the VEV alignments with the transformations of $A_4$.
We show that the 24 vacua are classified into two types in the following subsection.
Then, one finds the two different PMNS matrices with diagonalizing matrices for the charged lepton and effective neutrino mass matrices Eqs.~(\ref{eq:Ml1})-(\ref{eq:Ml4}), and (\ref{eq:Mv1}),(\ref{eq:Mv2}),(\ref{eq:Mv3}).
\subsection{Relations among Sets of VEV Alignments}
\begin{figure}[h]
	\begin{center}
	\includegraphics[width=7.0cm]{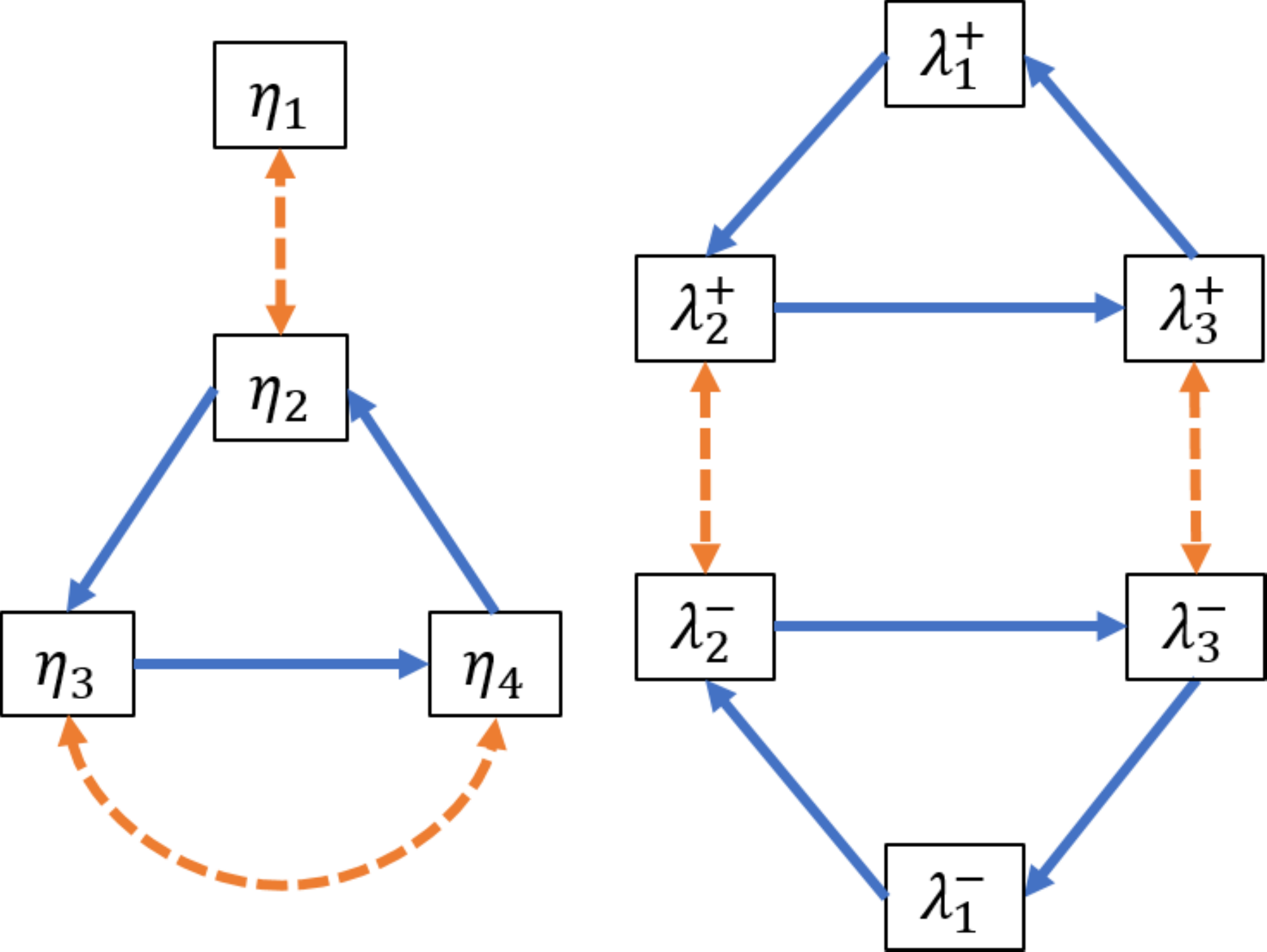}
	\caption{
		Map of the transitions among the VEV alignments under the transformations $S$ and $T$:
		The solid arrow corresponds to the transition due to $T$ transformation
		and the dashed two headed arrow shows the transition due to $S$ transformation.
		In the map, $ \eta_1$ is invariant under $T$ transformation
		while $\lambda_1^{\pm}$ are invariant under $S$ transformation. }
	\label{fig:map}
	\end{center}
\end{figure}

The generators of $A_4$ are expressed as the following forms for the representations ${\bf 1},{\bf 1'},{\bf 1''}$ and ${\bf 3}$,
\begin {align}
	&S({\bf 1})=S({\bf 1'})=S({\bf 1''})=1\ ,\quad S({\bf 3})
	=\frac{1}{3}\begin{pmatrix}
		-1&2&2\\
		2&-1&2\\
		2&2&-1
	\end{pmatrix},
\label{eq:Strans}\\
	&T({\bf 1})=1,\ T({\bf 1'})=\omega,\ T({\bf 1''})=\omega^2\ ,\quad T({\bf 3})=
	\begin{pmatrix}
		1&0&0\\
		0&\omega&0\\
		0&0&\omega^2
\label{eq:Ttrans}
	\end{pmatrix}.
\end{align}
The sets of VEV alignment $\eta_m,\lambda^\pm_n$ are associated through the transformations of these generators.
As an example, we show the $T$ transformation on $\lambda_1^+$:
\begin{align}
	T\left[\lambda_1^+\right]&\equiv
	\left\{\langle\phi_S\rangle=T({\bf 3})v_S
	\begin{pmatrix}
		1\\
		1\\
		1
	\end{pmatrix},
	\langle\xi'\rangle=T({\bf 1'})u',
	\langle\phi_0^S\rangle=T({\bf 3})
	\begin{pmatrix}
		0\\
		0\\
		0
	\end{pmatrix}\right\}\nonumber \\
	&=\left\{\langle\phi_S\rangle=v_S
	\begin{pmatrix}
		1\\
		\omega\\
		\omega^2
	\end{pmatrix},
	\langle\xi'\rangle=\omega u',\ 
	\langle\phi_0^S\rangle=
	\begin{pmatrix}
		0\\
		0\\
		0
	\end{pmatrix}\right\}=\lambda_2^+.
\end{align}
The $S$ and $T$ transformations on all the sets of the VEV alignment are summarized in Fig.~\ref{fig:map}.
Some transformations preserve the VEVs of either $\eta_m$ or $\lambda^\pm_n$.
These vacua have $Z_3$ or $Z_2$ symmetries as the residual symmetries of $A_4$ respectively.
For the VEVs described as $\eta_m$, they are invariant under the following transformation,
\begin{align}
	T\left[\eta_1\right]=T^{-1}\left[\eta_1\right]=\eta_1\ ,\quad
	TST\left[\eta_2\right]=(TST)^{-1}\left[\eta_2\right]=\eta_2 \ ,\nonumber\\
	ST\left[\eta_3\right]=(ST)^{-1}\left[\eta_3\right]=\eta_3\ ,\quad
	TS\left[\eta_4\right]=(TS)^{-1}\left[\eta_4\right]=\eta_4\ .
\end{align}
It is easy to confirm that such transformations correspond to $Z_3$ symmetries:
\begin{align}
	T^3=(TST)^3=(ST)^3=(TS)^3=1.
\end{align}
Each $\lambda^\pm_n$ has $Z_2$ symmetry as follows:
\begin{align}
	S\left[\lambda^\pm_1\right]=\lambda^\pm_1\ ,\quad TST^2\left[\lambda^\pm_2\right]
	=\lambda^\pm_2\ ,\quad T^2ST\left[\lambda^\pm_3\right]=\lambda^\pm_3,
\end{align}
where
\begin{align}
	S^2=(TST^2)^2=(T^2ST)^2=1.
\end{align}
\subsection{Classification of 24 Vacua}
In this subsection, we show the relations among the 24 different Lagrangians derived from the 24 different combinations of VEV alignments in Eqs.~\eqref{eq:eta1}-\eqref{eq:lam3}.
We find the two sets of 12 equivalent Lagrangians with the appropriate field redefinitions.
Then, the 24 Lagrangians are classified into two types.
For simplicity, we write the Lagrangian of this model in a short form:
\begin{align}
	\mathcal{L}(\psi,\phi_1,\phi_2)\label{eq:L},
\end{align}
where $\psi$ represents the fermion fields such as $l$ and $\nu_R$.
$\phi_1$ and $\phi_2$ represent the scalar fields,
which should have their VEVs written as $\eta_m$ and $\lambda_n^\pm$ respectively.
We write the Lagrangian in the broken phase for the VEV alignment ($\eta_m$, $\lambda^\pm_n$) with fluctuations $h_1$ and $h_2$ as
\begin{align}
	\mathcal{L}_{mn}^\pm(\psi,h_1,h_2)\equiv\mathcal{L}(\psi,\eta_m+h_1,\lambda_n^\pm+h_2).
\end{align}
Then, we prove the following equation:
\begin{align}
	\mathcal{L}(\psi',\eta_m+h_1',\lambda^\pm_n+h_2')=
	\mathcal{L}(\psi,G^{-1}\eta_m+h_1,G^{-1}\lambda_n^\pm+h_2),\label{eq:eqre}
\end{align}
where $G$ denotes the transformation composed of $S$ and $T$ in Eqs.~\eqref{eq:Strans} and \eqref{eq:Ttrans}.
There are 12 independent transformations including the identity element:
\begin{align}
	G;\{e,\ T,\ T^2,\ S,\ TS,\ T^2S,\ ST,\ ST^2,\ T^2ST,\ TST,\ TST^2,\ T^2ST^2\}.
	\label{eq:allG}
\end{align}
The redefined fields are written as follows,
\begin{align}
	\psi'=G\psi\ ,\quad h_i'=Gh_i\quad(i=1,2).\label{eq:redef}
\end{align}
The right-hand side of Eq.~\eqref{eq:eqre} corresponds to the Lagrangian for the vacuum ($G^{-1}\eta_m,G^{-1}\lambda^\pm_n$) while the left-hand side is the Lagrangian for the vacuum ($\eta_m,\lambda_n^\pm$) in terms of the redefined fields.
In the symmetric phase, the Lagrangian $\mathcal{L}(\psi,\phi_1,\phi_2)$ is invariant under the $G$ transformation,
\begin{align}
	\mathcal{L}(G\psi,G\phi_1,G\phi_2)=\mathcal{L}(\psi,\phi_1,\phi_2).\label{eq:INV}
\end{align}
One obtains the following equation from Eq.~\eqref{eq:INV} for the vacuum ($G^{-1}\eta_m,G^{-1}\lambda^\pm_n$),
\begin{align}
	\mathcal{L}(G\psi,\eta_m+Gh_1,\lambda_n^\pm+Gh_2)
	=&\mathcal{L}(\psi,G^{-1}\eta_m+h_1,G^{-1}\lambda_n^\pm+h_2).
\label{eq:aeqre}
\end{align}
Finally, one obtains the relation Eq.~\eqref{eq:eqre} by applying the field definition Eq.~\eqref{eq:redef} to the left-hand side of Eq.~\eqref{eq:aeqre}.
The relation Eq.~\eqref{eq:eqre} implies the equality of the Lagrangians for the two vacua ($\eta_m,\lambda^\pm_n$) and ($G^{-1}\eta_m,G^{-1}\lambda^\pm_n$).

Here, we briefly show how to find the equivalent vacua with Fig.~\ref{fig:map}.
For example, let us consider the $T$ transformation in terms of the vacuum of $(\eta_1,\lambda^+_1)$.
One finds that $\eta_1$ is invariant and $\lambda^+_1$ transfers to $\lambda^+_2$ under the $T$ transformation.
Therefore, $\mathcal{L}^+_{11}$ and $\mathcal{L}^+_{12}$ are equivalent.
One can find 12 equivalent vacua by applying 12 independent transformations in Eq.~\eqref{eq:allG} to the vacuum $(\eta_1,\lambda^+_1)$.
Then, we classify the 24 Lagrangians into two types:
\begin{align}
	&\mathrm{Type\ I};
	\{\mathcal{L}^+_{11},\mathcal{L}^+_{12},\mathcal{L}^+_{13},\mathcal{L}^+_{21},\mathcal{L}^+_{32},\mathcal{L}^+_{43},
	\mathcal{L}^-_{22},\mathcal{L}^-_{23},\mathcal{L}_{31}^-,\mathcal{L}^-_{33},\mathcal{L}^-_{41},\mathcal{L}^-_{42}\},\\
	&\mathrm{Type\ I\hspace{-.1em}I};
	\{\mathcal{L}^-_{11},\mathcal{L}^-_{12},\mathcal{L}^-_{13},\mathcal{L}^-_{21},\mathcal{L}^-_{32},\mathcal{L}^-_{43},
	\mathcal{L}^+_{22},\mathcal{L}^+_{23},\mathcal{L}^+_{31},\mathcal{L}^+_{33},\mathcal{L}^+_{41},\mathcal{L}^+_{42}\}.
\end{align}
Type I and type I\hspace{-.1em}I are disconnected because of the absence of a transformation which relates one type to the other.
Since all the Lagrangians which belong to the same type lead to the same physical consequences,
we consider only $\mathcal{L}^+_{11}$ and $\mathcal{L}^-_{11}$ as the representatives of their types:
\begin{align}
	\mathcal{L}^\mathrm{I}\equiv\mathcal{L}^+_{11}\ ,\ \ \mathcal{L}^{\mathrm{I\hspace{-.1em}I}}\equiv\mathcal{L}^-_{11}.
\end{align}
We also define the representative mass matrices for charged leptons and neutrinos as
\begin{align}
	M_l\equiv M_l^{(1)}\ ,\quad
	M_\nu^\mathrm{I}\equiv {M_\nu^{(1)}}^+\ ,\quad
	M_\nu^\mathrm{I\hspace{-.1em}I}\equiv {M_\nu^{(1)}}^-.
\end{align}
It is noted that the charged lepton mass matrix $M_l^{(1)}$ is diagonal.
\subsection{PMNS Matrices for Two Types}\label{sec:PMNS}
In this subsection, we construct the PMNS matrices for the two types,
$\mathcal{L}^\mathrm{I}$ and $\mathcal{L}^{\mathrm{I\hspace{-.1em}I}}$.
Since  the charged lepton mass matrix $M_l$ is diagonal,
the PMNS matrix is determined so that it diagonalizes the neutrino mass matrices in Eq.~(\ref{eq:Mv1}):
\begin{align}
	(U_\mathrm{PMNS}^\mathrm{I})^\dag M_\nu^\mathrm{I}(U_\mathrm{PMNS}^\mathrm{I})^*=
	(U_\mathrm{PMNS}^\mathrm{I\hspace{-.1em}I})^\dag M_\nu^\mathrm{I\hspace{-.1em}I}
	(U_\mathrm{PMNS}^\mathrm{I\hspace{-.1em}I})^*=
	\begin{pmatrix}
		m_1&&\\
		&m_2&\\
		&&m_3
	\end{pmatrix},
\label{eq:diag. cond.}
\end{align}
where the left-handed neutrino masses $m_1,m_2$ and $m_3$ are positive.
The PMNS matrices are expressed as the following forms for the two types:
\begin{align}
	U_\mathrm{PMNS}^\mathrm{I}&=U_\mathrm{TBM}U_{13}(\theta,\sigma)
	\begin{pmatrix}
		e^{i\phi_1}&& \\
		&e^{i\phi_2}& \\
		&&e^{i\phi_3}
	\end{pmatrix},
\label{eq:PMNSI}\\
	U_\mathrm{PMNS}^\mathrm{I\hspace{-.1em}I}&=U_\mathrm{TBM}\begin{pmatrix}
		&&-i\\
		&1&\\
		i&&
	\end{pmatrix}U_{13}(\theta,\sigma)\begin{pmatrix}
		e^{i\phi_1}&& \\
		&e^{i\phi_2}& \\
		&&e^{i\phi_3}
	\end{pmatrix}, \nonumber\\
	&=U_\mathrm{TBM}U_{13}^*(\theta+\frac{\pi}{2},\sigma)\begin{pmatrix}
		-ie^{i(\phi_1+\sigma)}&& \\
		&e^{i\phi_2}& \\
		&&-ie^{i(\phi_3-\sigma)}
	\end{pmatrix}.
\label{eq:PMNSII}
\end{align}
The unitary matrix $U_\mathrm{TBM}$ is the tri-bimaximal mixing matrix and $U_{13}(\theta,\sigma)$ denotes the unitary rotation matrix:
\begin{align}
	&U_\mathrm{TBM}=
	\begin{pmatrix}
		2/\sqrt{6}&1/\sqrt{3}&0\\
		-1/\sqrt{6}&1/\sqrt{3}&-1/\sqrt{2}\\
		-1/\sqrt{6}&1/\sqrt{3}&1/\sqrt{2}
	\end{pmatrix},\label{eq:TBM}\\
	&U_{13}(\theta,\sigma)=
	\begin{pmatrix}
		\mathrm{cos}\theta&0&e^{-i\sigma}\mathrm{sin}\theta\\
		0&1&0\\
		-e^{i\sigma}\mathrm{sin}\theta&0&\mathrm{cos}\theta
	\end{pmatrix}.\label{eq:R13}
\end{align}
We have introduced the parameters $\theta,\sigma$ and $\phi_i$ ($i=1,2,3$).
They are written in terms of the complex parameters of the neutrino mass matrix, $a,b,c$ and $d$, in Eq.~\eqref{eq:Mv1}~\footnote{
	There are six real parameters since $b$ is written by using $a,c,d$.}.
In the rest of this subsection, we derive the explicit forms of the parameters $\theta,\sigma$ and $\phi_i$ in terms of the model parameters $a,b,c$ and $d$.
In the first step, one rotates $M_\nu M_\nu^\dag$ with the tri-bimaximal mixing matrix.
\begin{align}
	&U_\mathrm{TBM}^\dag M_\nu^\mathrm{I}(M_\nu^\mathrm{I})^\dag U_\mathrm{TBM}=
	\begin{pmatrix}
		A&0&B\\
		0&C&0\\
		B^*&0&D
	\end{pmatrix},
\label{eq:MMdagTBM1}\\
	&U_\mathrm{TBM}^\dag M_\nu^\mathrm{I\hspace{-.1em}I}(M_\nu^\mathrm{I\hspace{-.1em}I})^\dag U_\mathrm{TBM}=
	\begin{pmatrix}
		D&0&-B^*\\
		0&C&0\\
		-B&0&A
	\end{pmatrix}.
\label{eq:MMdagTBM2}
\end{align}
where
\begin{align}
	&A=\left|a+c-\frac{d}{2}\right|^2+\left|\frac{\sqrt{3}}{2}d\right|^2, \label{eq:A} \\
	&B=\left(a+c-\frac{d}{2}\right)\frac{\sqrt{3}}{2}d^*+\frac{\sqrt{3}}{2}d\left(a-c+\frac{d}{2}\right)^* \label{eq:B}
	\equiv|B|e^{i\varphi_B}, \\
	&C=\left|\frac{a^2-(c^2-cd+d^2)}{2d-c}\right|^2, \label{eq:C} \\
	&D=\left|a-c+\frac{d}{2}\right|^2+\left|\frac{\sqrt{3}}{2}d\right|^2. \label{eq:D}
\end{align}
The mass eigenvalues are determined as
\begin{align}
	m_1^2&=\frac{A+D}{2}\mp\frac{1}{2}\sqrt{(A-D)^2+4|B|^2},\\
	m_2^2&=C,\\
	m_3^2&=\frac{A+D}{2}\pm\frac{1}{2}\sqrt{(A-D)^2+4|B|^2},
\end{align}
where the upper and lower signs in these mass eigenvalues correspond to the normal hierarchy (NH) and the inverted hierarchy (IH).
Next, we diagonalize the rotated mass matrices, Eqs.~\eqref{eq:MMdagTBM1} and \eqref{eq:MMdagTBM2},
with $U_{13}(\theta,\sigma)$ and $U_{13}^*(\theta+\frac{\pi}{2},\sigma)$ respectively:
\begin{align}
	&U_{13}(\theta,\sigma)^\dag
	\begin{pmatrix}
		A&0&B\\
		0&C&0\\
		B^*&0&D
	\end{pmatrix}
	U_{13}(\theta,\sigma)
	=\begin{pmatrix}
		m_1^2&&\\
		&m_2^2&\\
		&&m_3^2
	\end{pmatrix}, \\
	&U_{13}(\theta+\frac{\pi}{2},\sigma)^\mathrm{T}
	\begin{pmatrix}
		D&0&-B^*\\
		0&C&0\\
		-B&0&A
	\end{pmatrix}
	U_{13}(\theta+\frac{\pi}{2},\sigma)^*
	=\begin{pmatrix}
		m_1^2&&\\
		&m_2^2&\\
		&&m_3^2
	\end{pmatrix},
\end{align}
where $\theta$ and $\sigma$ are determined as,
\begin{align}
	\mathrm{tan}2\theta=\frac{2|B|}{D-A}\ ,\ \sigma=-\varphi_B.
\end{align}
Finally, the other parameters $\phi_i$ are determined as follows,
\begin{align}
	&\phi_1=\nonumber\\
	&\frac{1}{2}\left[\mathrm{tan}^{-1}\left[\frac
	{\left(\mathrm{Im}\left[a\right]+\mathrm{Im}\left[c-\frac{d}{2}\right]\mathrm{cos}2\theta\right)\mathrm{cos}\sigma
	+\left(\mathrm{Re}\left[a\right]\mathrm{cos}2\theta+\mathrm{Re}\left[c-\frac{d}{2}\right]\right)\mathrm{sin}\sigma
	-\frac{\sqrt{3}}{2}\mathrm{Im}\left[d\right]\mathrm{sin}2\theta}
	{\left(\mathrm{Re}\left[a\right]+\mathrm{Re}\left[c-\frac{d}{2}\right]\mathrm{cos}2\theta\right)\mathrm{cos}\sigma
	-\left(\mathrm{Im}\left[a\right]\mathrm{cos}2\theta+\mathrm{Im}\left[c-\frac{d}{2}\right]\right)\mathrm{sin}\sigma
	-\frac{\sqrt{3}}{2}\mathrm{Re}\left[d\right]\mathrm{sin}2\theta}\right]-\sigma\right],\\
	&\phi_2=\frac{1}{2}\mathrm{tan}^{-1}\left[\frac
	{\mathrm{Im}\left[a^2-(c^2-cd+d^2)\right]\mathrm{Re}\left[2d-c\right]
	-\mathrm{Re}\left[a^2-(c^2-cd+d^2)\right]\mathrm{Im}\left[2d-c\right]}
	{\mathrm{Re}\left[a^2-(c^2-cd+d^2)\right]\mathrm{Re}\left[2d-c\right]
	+\mathrm{Im}\left[a^2-(c^2-cd+d^2)\right]\mathrm{Im}\left[2d-c\right]}\right],\\
	&\phi_3=\nonumber\\&\frac{1}{2}\left[\mathrm{tan}^{-1}\left[\frac
	{\left(\mathrm{Im}\left[a\right]-\mathrm{Im}\left[c-\frac{d}{2}\right]\mathrm{cos}2\theta\right)\mathrm{cos}\sigma
	-\left(\mathrm{Re}\left[a\right]\mathrm{cos}2\theta-\mathrm{Re}\left[c-\frac{d}{2}\right]\right)\mathrm{sin}\sigma
	+\frac{\sqrt{3}}{2}\mathrm{Im}\left[d\right]\mathrm{sin}2\theta}
	{\left(\mathrm{Re}\left[a\right]-\mathrm{Re}\left[c-\frac{d}{2}\right]\mathrm{cos}2\theta\right)\mathrm{cos}\sigma
	+\left(\mathrm{Im}\left[a\right]\mathrm{cos}2\theta-\mathrm{Im}\left[c-\frac{d}{2}\right]\right)\mathrm{sin}\sigma
	+\frac{\sqrt{3}}{2}\mathrm{Re}\left[d\right]\mathrm{sin}2\theta}\right]+\sigma\right].
\end{align}
We briefly explain the derivation of $\phi_i$ for the mass matrix $M_\nu^{\mathrm{I}}$.
We first diagonalize $M_\nu^{\mathrm{I}}$ with the unitary matrices $U_\mathrm{TBM}$ and $U_{13}(\theta,\sigma)$ according to Eq.~\eqref{eq:diag. cond.}.
However, the diagonalized neutrino mass matrix consists of complex elements.
Then, the parameters $\phi_i$ are determined so that all the elements of the diagonalized matrix are real and positive.
\section{Phenomenological Aspects}\label{sec:num}
We study the phenomenological aspects of this model and show the differences between the two types of vacua.
The observables, such as mixing angles and CP violating phases, are described with the model parameters in different forms for the two types.
In the following subsections, we discuss the relation between the observables and model parameters.
The numerical analyses are also shown in this section.
\subsection{Mixing Angles and CP Violating Phases}
In this subsection, we discuss the lepton mixing angles, CP violating phases and the effective mass for $0\nu\beta\beta$ decay.
At first, we introduce the PDG parametrization of the PMNS matrix:
\begin{align}
	U_{\mathrm{PMNS}}^{\mathrm{PDG}}=
	\begin{pmatrix}
		c_{12}c_{13} & s_{12}c_{13} & s_{13}e^{-i\delta_{CP}} \\
		-s_{12}c_{23}-c_{12}s_{23}s_{13}e^{i\delta_{CP}} & c_{12}c_{23}-s_{12}s_{23}s_{13}e^{i\delta_{CP}} & s_{23}c_{13} \\
		s_{12}c_{23}-c_{12}c_{23}s_{13}e^{i\delta_{CP}} & -c_{12}s_{23}-s_{12}c_{23}s_{13}e^{i\delta_{CP}} & c_{23}c_{13}
	\end{pmatrix}
	\begin{pmatrix}
		e^{i\alpha}&&\\
		&e^{i\beta}&\\
		&&1
	\end{pmatrix},
\label{eq:PDG}
\end{align}
where $s_{ij}$ and $c_{ij}$ denote the lepton mixing angles $\sin\theta_{ij}$ and $\cos\theta_{ij}$, respectively.
They are written in terms of the PMNS matrix elements:
\begin{align}
	\sin^2\theta_{12}=\frac{|U_{e2}|^2}{1-|U_{e3}|^2}\ ,\quad
	\sin^2\theta_{23}=\frac{|U_{\mu3}|^2}{1-|U_{e3}|^2}\ ,\quad
	\sin^2\theta_{13}=|U_{e3}|^2,
\label{eq:moduli}
\end{align}
where $U_{\alpha i}$ denote the PMNS matrix elements.
The Dirac CP violating phase $\delta_{CP}$ can be obtained with the Jarlskog invariant
\begin{align}
	\sin\delta_{CP}&=\frac{J_{CP}}{s_{23}c_{23}s_{12}c_{12}s_{13}c_{13}^2},\label{eq:cp}\\
	J_{CP}&=\mathrm{Im}\left[U_{e1}U_{\mu2}U_{\mu1}^*U_{e2}^*\right].
\label{eq:jcp}
\end{align}
In order to obtain these parameters from our model, we substitute the PMNS matrix elements in Eqs.~\eqref{eq:PMNSI} and \eqref{eq:PMNSII}.
For the type I case, the matrix elements are given as follows:
\begin{align}
	U_{e1}&=\frac{2}{\sqrt{6}}e^{i\phi_1}\cos\theta,\\
	U_{e2}&=U_{\mu2}=\frac{1}{\sqrt{3}} e^{i\phi_2},\\
	U_{e3}&=\frac{2}{\sqrt{6}}e^{-i(\sigma-\phi_3)}\sin\theta,\\
	U_{\mu1}&=\left(-\frac{1}{\sqrt{6}}\cos\theta+\frac{1}{\sqrt{2}}e^{i\sigma}\sin\theta\right)e^{i\phi_1},\\
	U_{\mu3}&=\left(-\frac{1}{\sqrt{6}}e^{-i\sigma}\sin\theta-\frac{1}{\sqrt{2}}\cos\theta\right)e^{i\phi_3}.
\end{align}
The mixing angles, Dirac CP violating phase and Majorana phases for both types are listed in Table~\ref{tb:angles}.
\begin{table}[h]
	\caption{Mixing angles, Dirac CP phase and Majorana phases for the two types of vacua}
	\centering
	\begin{tabular}{|c||c|c|} \hline
		&$\mathrm{Type~I}$&$\mathrm{Type~I\hspace{-.1em}I}$\\ \hline \hline

		$\mathrm{sin}^2\theta_{12}$&
		${\displaystyle \frac{1}{2+\mathrm{cos}2\theta}}$&
		${\displaystyle \frac{1}{2-\mathrm{cos}2\theta}}$\\[0.3cm]

		$\mathrm{sin}^2\theta_{23}$&
		${\displaystyle \frac{1}{2}(1+\frac{\sqrt{3}\mathrm{sin}2\theta}{2+\mathrm{cos}2\theta}\mathrm{cos}\sigma)}$&
		${\displaystyle \frac{1}{2}(1-\frac{\sqrt{3}\mathrm{sin}2\theta}{2-\mathrm{cos}2\theta}\mathrm{cos}\sigma)}$\\[0.3cm]

		$\mathrm{sin}^2\theta_{13}$&
		${\displaystyle \frac{1}{3}(1-\mathrm{cos}2\theta)}$&
		${\displaystyle \frac{1}{3}(1+\mathrm{cos}2\theta)}$\\[0.3cm]

		$\mathrm{sin}\delta_\mathrm{CP}$&
		${\displaystyle -\frac{\sin2\theta}{|\sin2\theta|}\frac{(2+\mathrm{cos}2\theta)\mathrm{sin}\sigma}
		{\sqrt{(2+\mathrm{cos}2\theta)^2-3\mathrm{sin}^22\theta\mathrm{cos}^2\sigma}}}$&
		${\displaystyle -\frac{\sin2\theta}{|\sin2\theta|}\frac{(2-\mathrm{cos}2\theta)\mathrm{sin}\sigma}
		{\sqrt{(2-\mathrm{cos}2\theta)^2-3\mathrm{sin}^22\theta\mathrm{cos}^2\sigma}}}$ \\[0.3cm]

		$\alpha+\delta_{\mathrm{CP}}$ & $\phi_1-\phi_3+\sigma$ & $\phi_1-\phi_3+\sigma$ \\
		$\beta+\delta_{\mathrm{CP}}$ & $\phi_2-\phi_3+\sigma$ &  ${\displaystyle \phi_2-\phi_3+\frac{\pi}{2}}$ \\ \hline
	\end{tabular}
\label{tb:angles}
\end{table}
One can adopt either of the two types to predict the mixing angles and the Dirac CP violating phases,
since  both types give the same predictions.
However, we note the following two facts.
First, if one fixes $\cos2\theta\simeq1$ to obtain small $\sin^2\theta_{13}$ in type I,
$\sin^2\theta_{13}$ in type I\hspace{-.1em}I reaches $2/3$, which is disfavored in the experiments.
Second, as shown in Subsection \ref{sec:PMNS}, the model parameters $\theta$, $\sigma$ and $\phi_i$ are expressed in the same forms for the two types with $a$, $b$, $c$ and $d$ of Eq.~\eqref{eq:Mv1}.
Therefore, those parameters have common values for both types.
Hence, the two types can not realize the experimental results simultaneously.

Next, we discuss the effective mass for $0\nu\beta\beta$ decay, $m_{ee}=\sum_im_iU_{ei}^2$,
and the Majorana phases, $\alpha$ and $\beta$.
The effective mass is given as follows:
\begin{align}
	\left|m_{ee}^\mathrm{I}\right|&=
	\frac{1}{3}\left|m_1(1+\mathrm{cos}2\theta)e^{2i\phi_1}
	+m_2e^{2i\phi_2}+m_3(1-\mathrm{cos}2\theta)e^{2i(\phi_3-\sigma)}\right|,
\label{eq:Mee1}\\
	\left|m_{ee}^\mathrm{I\hspace{-.1em}I}\right|&=
	\frac{1}{3}\left|m_1(1-\mathrm{cos}2\theta)e^{2i(\phi_1+\sigma)}
	-m_2e^{2i\phi_2}+m_3(1+\mathrm{cos}2\theta)e^{2i\phi_3}\right|,
\label{eq:Mee2}
\end{align}
where the superscripts I and I\hspace{-.1em}I denote the types of vacuum.
Equivalently, one can rewrite Eqs.~\eqref{eq:Mee1} and \eqref{eq:Mee2} as
\begin{align}
	\left|m_{ee}^\mathrm{I}\right|&=
	\frac{1}{3}\left|m_1(1+\mathrm{cos}2\theta)e^{2i(\phi_1-\phi_3+\sigma)}
	+m_2e^{2i(\phi_2-\phi_3+\sigma)}+m_3(1-\mathrm{cos}2\theta)\right|,
\label{eq:Mee3}\\
	\left|m_{ee}^\mathrm{I\hspace{-.1em}I}\right|&=
	\frac{1}{3}\left|m_1(1-\mathrm{cos}2\theta)e^{2i(\phi_1-\phi_3+\sigma)}
	-m_2e^{2i(\phi_2-\phi_3)}+m_3(1+\mathrm{cos}2\theta)\right|,
\label{eq:Mee4}
\end{align}
On the other hand, the effective mass in the PDG parametrization is written as
\begin{align}
	\left|m_{ee}\right|=\left|
	m_1c_{13}^2c_{12}^2e^{2i(\alpha+\delta_{CP})}+
	m_2c_{13}^2s_{12}^2e^{2i(\beta+\delta_{CP})}+
	m_3s_{13}^2\right|.
\label{eq:MeePDG}
\end{align}
One can obtain the Majorana CP violating phases $\alpha$ and $\beta$ by comparing Eqs~\eqref{eq:Mee3}-\eqref{eq:MeePDG},
\begin{align}
	(\mathrm{Type~I})\quad
	&\alpha+\delta_{CP}=\phi_1-\phi_3+\sigma\ ,\quad
	\beta+\delta_{CP}=\phi_2-\phi_3+\sigma,
\label{eq:phase1}\\[0.3cm] 
	(\mathrm{Type~I\hspace{-.1em}I})\quad
	&\alpha+\delta_{CP}=\phi_1-\phi_3+\sigma\ ,\quad
	\beta+\delta_{CP}=\phi_2-\phi_3+\frac{\pi}{2}.
\label{eq:phase2}
\end{align}
\subsection{Numerical Analysis}
In this subsection, we show numerical analysis to find the difference between two types.
We use recent experimental results with $3\sigma$ range~\cite{Esteban:2016qun}, as summarized in Table~\ref{tb:exp}.
\begin{table}[h]
	\caption{
		The experimental data for the mass squared differences and mixing angles with $3\sigma$ range~\cite{Esteban:2016qun}}
	\centering
	\begin{tabular}{|c||c|c|} \hline
		& Normal Hierarchy (NH) & Inverted Hierarchy (IH) \\ \hline
		$\Delta m_{21}^2~\left[\mathrm{eV}^2\right]$ & $(7.03\sim8.09)\times10^{-5}$ & $(7.03\sim8.09)\times10^{-5}$ \\
		$\Delta m_{31}^2~\left[\mathrm{eV}^2\right]$ & $(2.407\sim2.643)\times10^{-3}$ & $-(2.565\sim2.318)\times10^{-3}$ \\
		$\mathrm{sin}^2\theta_{12}$ & $0.271\sim0.345$ & $0.271\sim0.345$ \\
		$\mathrm{sin}^2\theta_{23}$ & $0.385\sim0.635$ & $0.393\sim0.640$ \\
		$\mathrm{sin}^2\theta_{13}$ & $0.01934\sim0.02392$ & $0.01953\sim0.02408$ \\ \hline
	 \end{tabular}
\label{tb:exp}
\end{table}
As we have shown in the previous subsection, the mixing angles and Dirac CP phase are expressed in terms of the model parameters $\theta$ and $\sigma$ in different forms for the two types.

The experimental data for $\sin^2\theta_{13}$ in Table~\ref{tb:exp} is realized by the following value of $\theta$ with NH or IH:
\begin{align}
	\mathrm{Type~I}\ ;\quad
	9.81^\circ\leq |\theta| \leq10.9^\circ\ (\mathrm{NH})\ ,\ 
	9.86^\circ\leq |\theta| \leq11.0^\circ\ (\mathrm{IH}),\\
	\mathrm{Type~I\hspace{-.1em}I}\ ;\quad
	79.1^\circ\leq |\theta| \leq80.2^\circ\ (\mathrm{NH})\ ,\ 
	79.0^\circ\leq |\theta| \leq80.1^\circ\ (\mathrm{IH}).
\end{align}
The value of $\sigma$ is allowed in  $-180^\circ\leq\sigma\leq180^\circ$ for both of the two types,
since the error of $\sin^2\theta_{23}$ from the experiments is large.

\begin{figure}[h]
	\begin{center}
	\includegraphics[width=7.0cm]{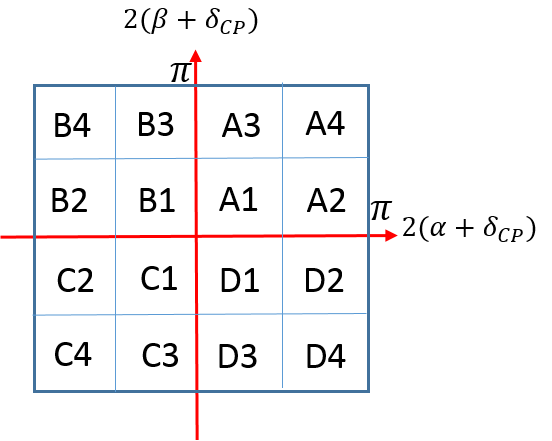}
 	\caption{16 divided regions for $2(\alpha+\delta_{CP})$ and $2(\beta+\delta_{CP})$.}
	\label{fig:block}
	\end{center}
\end{figure}

\begin{figure}[h]
	\begin{center}
	\includegraphics[width=0.7\linewidth]{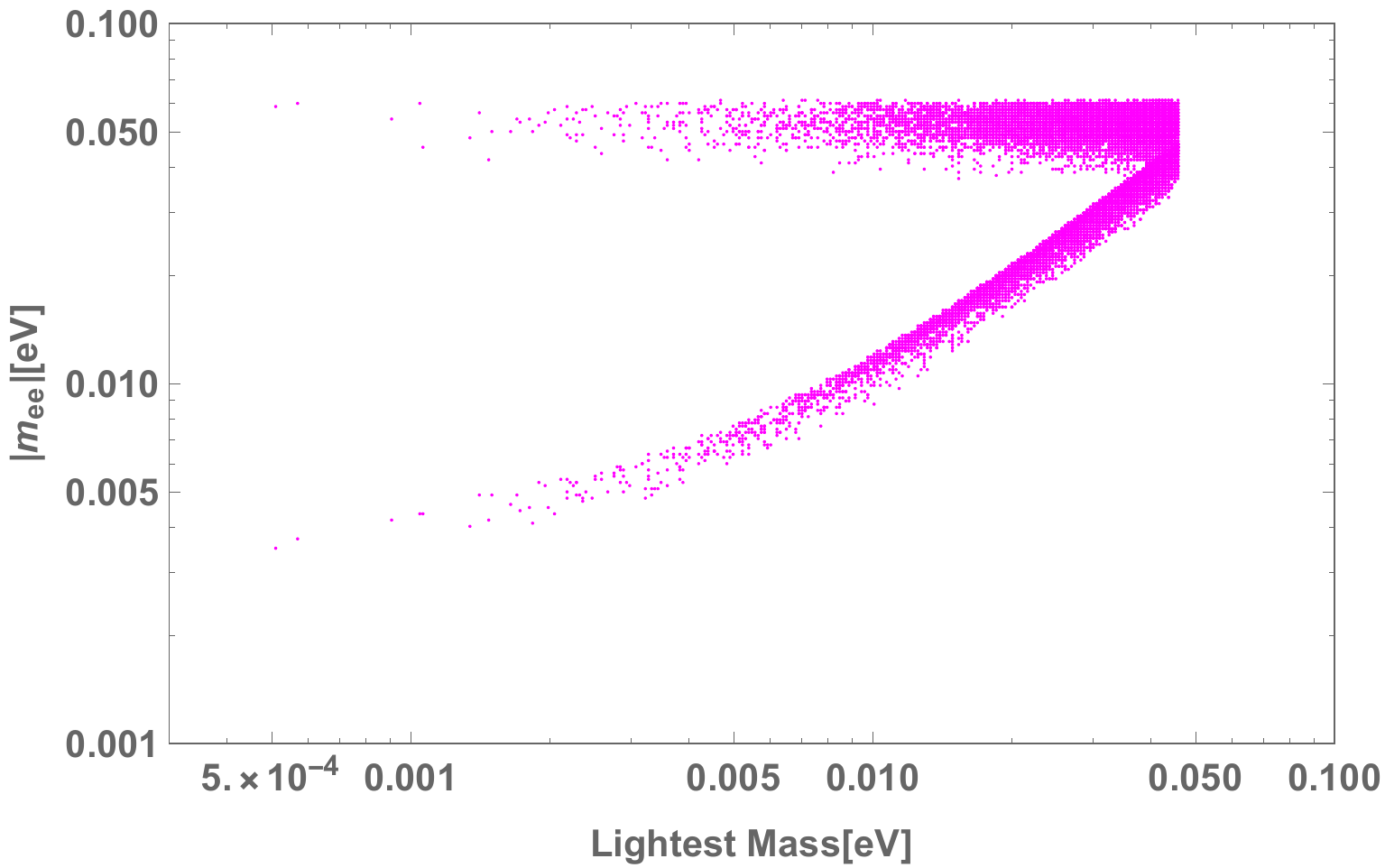}
	\caption{
		The prediction of effective mass for $0\nu\beta\beta$ decay in region A1 of Fig.~\ref{fig:block}.    
		The upper region corresponds to the IH case , while the lower one corresponds to the NH case.}
	\label{fig:meeA1}
	\end{center}
\end{figure}

\begin{figure}[h!]
	\begin{tabular}{cc}
		\begin{minipage}{0.5\hsize}
		\begin{center}
		\includegraphics[width=\linewidth]{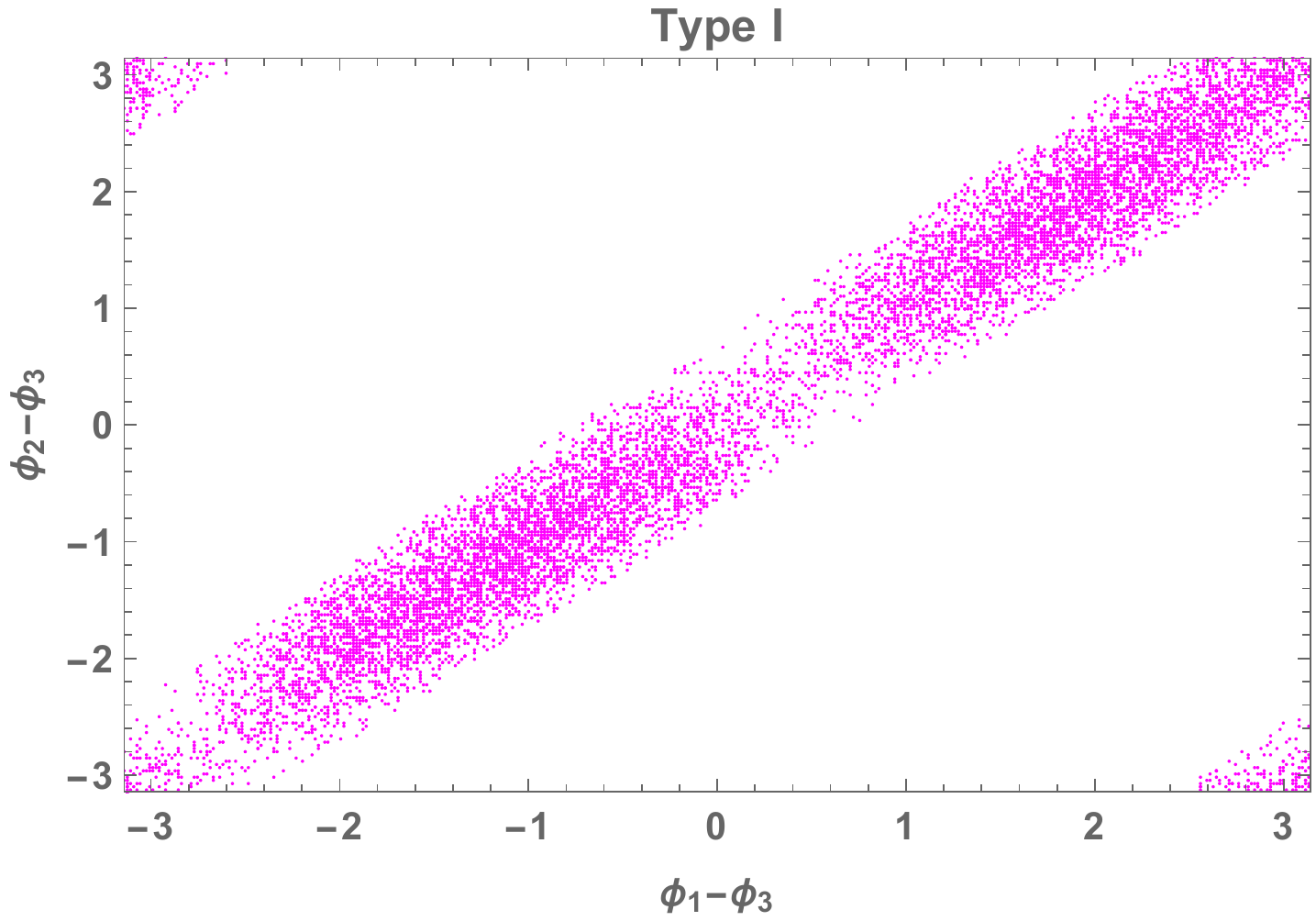}
		\end{center}
		\end{minipage}
		\begin{minipage}{0.5\hsize}
		\begin{center}
		\includegraphics[width=\linewidth]{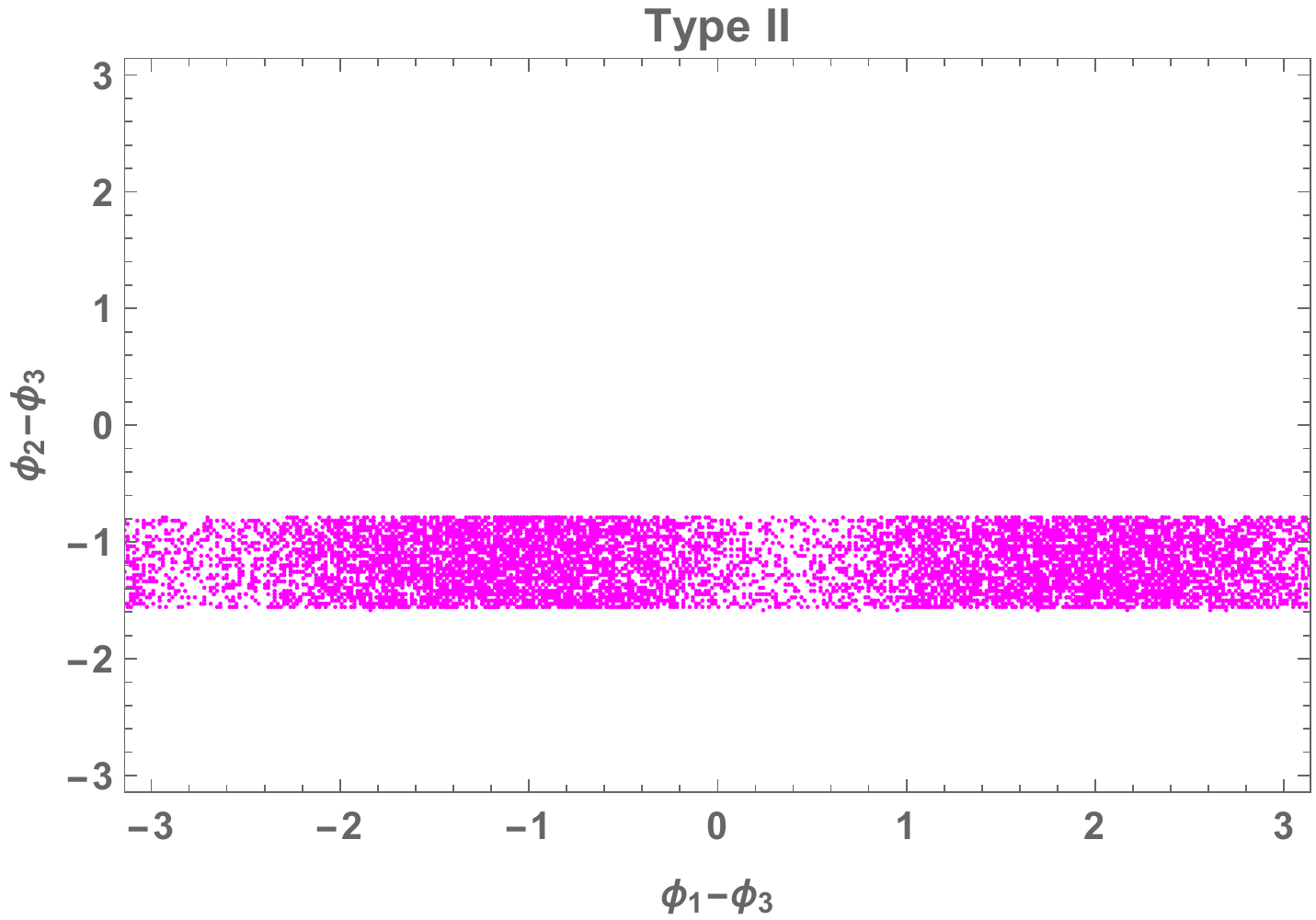}
		\end{center}
		\end{minipage}
	\end{tabular}
	\caption{
		The allowed regions of the model parameters, $\phi_1-\phi_3$ and $\phi_2-\phi_3$ for both types of vacua.
		These plots correspond to region A1 of Fig.~\ref{fig:block}.}
	\label{fig:1323}
\end{figure}

Next, we discuss the parameters $\phi_i$ in the expressions of the Majorana phases of Eqs.~\eqref{eq:phase1} and \eqref{eq:phase2}.
The effective mass $|m_{ee}|$ in Eq.~\eqref{eq:MeePDG} depends on the two combinations of Dirac and Majorana phases, $2(\alpha+\delta_{CP})$ and $2(\beta+\delta_{CP})$.
If we determine both $|m_{ee}|$ and the lightest neutrino mass, we obtain the constraints on these two combinations.
In order to find how the numerical constraints on $\phi_i$ are different in the two types,
we consider a specific situation.
As an example, we assume that $|m_{ee}|$ is predicted in the region as shown in Fig.~\ref{fig:meeA1}.
We note that the lightest neutrino mass is constrained from the cosmological upper bound for the neutrino mass sum, $\sum_im_i<0.16~\mathrm{eV}$~\cite{Alam:2016hwk}.
This plot is obtained when the Dirac and Majorana phases are randomly chosen from the region A1 in Fig.~\ref{fig:block},
\begin{align}
	0 < \alpha + \delta_{CP} < \pi/4\ ,\  0< \beta + \delta_{CP} < \pi/4.
\end{align}
In this situation, the phase differences $\phi_1 - \phi_3$ and $\phi_2 - \phi_3$ for one type can be distinguished from those for the other type.
The constraints on the phase differences are shown in Fig.~\ref{fig:1323}.
For type I, the phase difference $\phi_2-\phi_3$ is proportional to $\phi_1-\phi_3$.
However, for type I\hspace{-.1em}I, $\phi_2-\phi_3$ is independent of the value of $\phi_1-\phi_3$ because $\sigma$ is absent in the expression of $\phi_2-\phi_3$ in Eq.~\eqref{eq:phase2}.

\newpage
\section{Summary}\label{sec:con}

We have studied phenomenological aspects of a supersymmetric model with $A_4\times Z_3$ symmetry.
We found 24 degenerate vacua at the 24 minima of the scalar potential.
Then, we discussed the relations among the 24 different vacua and classified them into two types.
Both types consist of 12 vacua which are related to each other by transformations of $A_4$.
We proved that the 12 vacua are equivalent and lead to the same physical consequences.
However, we found that we obtain different physical consequences from the vacua of different types.
Therefore, we analyzed the two types of vacua to find the different phenomenological consequences of the two types.
In particular, we investigated observables such as mixing angles, Dirac CP phase, Majorana phases and effective mass for $0\nu\beta\beta$ decay.

These observables are expressed in terms of the model parameters $\theta$, $\sigma$ and $\phi_i$.
The angle $\theta$ and phase $\sigma$ are determined by the deviation from the tri-bimaximal mixing matrix.
The two types lead to different expressions for the mixing angles and Dirac CP violating phase in terms of $\theta$ and $\sigma$.
Therefore, one should take different model parameters in each type in order to realize the experimental results.
Although one can adopt both of the two types to predict the observable parameters,
the two types cannot realize the current experimental data simultaneously.
The Majorana phases $\alpha$ and $\beta$ are parametrized in the different expressions for each type by the model parameters $\phi_i$ in addition to $\theta$ and $\sigma$.
In order to find numerical differences between the two types of  Majorana phase,
we considered the specific situation where the lightest mass and effective mass for the $0\nu\beta\beta$ decay are determined in a certain region.
We showed the allowed regions of the phase differences, $\phi_1-\phi_3$ and $\phi_2-\phi_3$.
The regions are quite different for the two types:
the phase differences for type I are proportional to each other,
while those for type I\hspace{-.1em}I are not.

The VEVs $\eta_m$ and $\lambda^\pm_n$ transfer to the different VEVs by  transformations of $A_4$.
However,  the transformations for $\eta_m$ and $\lambda^\pm_n$ are closed differently since they have the $Z_3$ and $Z_2$ residual symmetries from $A_4$ respectively.
We have pointed out that some combinations of the VEVs can lead to different physical consequences.
When we consider models with two or more flavons, we should take account of the combination of VEVs.

\vspace{0.5cm}
\noindent
{\bf Acknowledgement}

\noindent
This work is supported by JSPS KAKENHI Grant Number JP17K05418 (T.M.).
This work is also supported in part by Grants-in-Aid for Scientific Research [No.~16J05332 (Y.S.), Nos. 24540272, 26247038, 15H01037, 16H00871, and 16H02189 (H.U.)] from the Ministry of Education,
Culture, Sports, Science and Technology in Japan.
H.O. is also supported by Hiroshima Univ. Alumni Association.

\newpage
\appendix
\section{Multiplication rule of $A_4$ group}
\label{sec:multiplication-rule}
In this appendix, we show the multiplication of the $A_4$ group.
The multiplication rule of the triplets is
written as follows;
\begin{align}
	\begin{pmatrix}
		a_1\\
		a_2\\
		a_3
	\end{pmatrix}_{\bf 3}
	\otimes
	\begin{pmatrix}
		b_1\\
		b_2\\
		b_3
	\end{pmatrix}_{\bf 3}
	&=\left (a_1b_1+a_2b_3+a_3b_2\right )_{\bf 1}
	\oplus \left (a_3b_3+a_1b_2+a_2b_1\right )_{{\bf 1}'} \nonumber \\
	& \oplus \left (a_2b_2+a_1b_3+a_3b_1\right )_{{\bf 1}''} \nonumber \\
	&\oplus \frac13
	\begin{pmatrix}
		2a_1b_1-a_2b_3-a_3b_2 \\
		2a_3b_3-a_1b_2-a_2b_1 \\
		2a_2b_2-a_1b_3-a_3b_1
	\end{pmatrix}_{{\bf 3}}
	\oplus \frac12
	\begin{pmatrix}
		a_2b_3-a_3b_2 \\
		a_1b_2-a_2b_1 \\
		a_3b_1-a_1b_3
	\end{pmatrix}_{{\bf 3}\ ,}
\end{align}
while that for singlets is,
\begin{align}
	{\bf 1}'\otimes {\bf 1}''={\bf 1}.
\end{align}
In order to derive the $A_4$ invariant superpotential in Eq.~\eqref{eq:wY}, we have used the multiplication rules.
Their derivation is shown in the reviews in Refs.~\cite{Ishimori:2010au,Ishimori:2012zz,Ishimori:2013woa,King:2014nza}.

\end{document}